\documentstyle[epsfig,longtable]{aipproc}
\begin{document}
\title{On Higgs Production in $\gamma \gamma$ Collisions}
\author{Oleg Yakovlev}
\address{Randall Laboratory of Physics, University of Michigan, 
Ann Arbor,  Michigan 48109-1120, USA}
\maketitle
\begin{abstract}
I review recent progress on the Higgs production in $\gamma \gamma$
collisions at the photon mode of the Next Linear Collider (NLC). 
I mainly focus on two particular topics. 
The first topic is the Higgs-two photon vertex, which is sensitive 
to new  physics, and can be considered a counter of 
the number of new heavy particles. I recall the results on QCD and 
electroweak two loop radiative corrections. 
The second topic is the heavy quark anti-quark pair production in $\gamma
\gamma$ collisions, which is the dominant background process for the 
Higgs production at $m_H <150$ GeV. We suggest a procedure for the resummation 
of double (DL) and single (SL) logarithms in the process 
$\gamma \gamma \to b \bar b$.   
\end{abstract}
\section{Introduction}
The neutral scalar Higgs boson \cite{Higgs} is an important ingredient of the
Standard Model (SM) and is the only SM elementary particle
which has not been detected so far (see for 
the review \cite{Kane}). A lower limit on $m_H$, of approximately $113.5$
GeV at $95\%  c.l.$,  has been obtained from direct searches at 
LEP \cite{LEPHIGGS}.
Current experiments  are concentrating on the possibility of finding a
Higgs particle in the intermediate mass region 
$113.5 < m_H < 150\quad\mbox{GeV}.$ In this region the Higgs particle
decays mainly to a $b\bar b$ pair. 

The {\it photon  mode} of the NLC, namely the collisions of the
energetic polarized Compton photons, 
will be used for the production and studies of the Higgs particle.

In this talk I review recent progress on the Higgs production in $\gamma \gamma$
collisions at the NLC. I will mainly focus on:\\ 
1) the Higgs-$\gamma \gamma$ vertex, which is sensitive to new physics, 
and can be considered the counter of the number of new heavy particles.
I review the results of QCD and electroweak radiative corrections performed in 
\cite{MeYa1,MeYa2,ZeSp1,KoYa,MeStHqq}.\\ 
2) the heavy quark anti-quark pair production in $\gamma \gamma$
collisions, which is main background of the Higgs production for
$m_H <150$ GeV. 
The radiative corrections to $\gamma \gamma \to b \bar b$ are extremely 
large \cite{Jikia} and are dominated by large QCD double logarithms (DL)
\cite{FKM}. We have resummed large QCD double logarithms of the form 
$(\alpha_s \ln^2 (s/m^2) )^n$ \cite{AkWaYa,MeSt1,KoYa}. We have 
derived the next-to-leading logarithmic correction to the DL result, 
which is effectively an resummation of all terms of the form  
$\alpha^n_s \ln^{2n-1} (s/m^2)$\cite{AkWaYa}. 
 \section{Higgs-two photon vertex}
The coupling of the Higgs boson with two photons is absent at 
tree level in the Standard Model. The first non-zero contribution 
arises from  fermions and W boson loops. Because the Yukawa coupling 
of the quark is proportional to the quark mass, 
the contributions of the light quarks as well as charm and 
bottom quarks are well suppressed in comparison to the top quark loop
contribution.  
Because we intend to use the Higgs-$\gamma\gamma$ coupling 
as a counter for new undiscovered particles, it should be a very well
calibrated tool.  One important question is: how large are QCD corrections?
The answer is: the QCD corrections are small for Higgs masses 
$m_H < 2m_t$ \cite{ZeSp1,MeYa1}. This has been shown by the explicit
calculations in  \cite{ZeSp1,MeYa1}. 
The only source of QCD corrections at two loop  order is the gluon
corrections to the top quark loop. 
For the heavy Higgs masses, $m_H > 2 m_t$, the corrections are large
(about 40\%), although this mass range is ruled out by the electro-weak
data. 
It was observed in \cite{MeYa1} that in the limit of large
ratio $\frac{m_H}{m_q}$  the form factor gives the QCD double 
logarithmic asymptotic, $(1-\frac{\rho}{12})$, 
with $\rho =\frac{C_F\alpha_s(\mu^2)}{2\pi}\ln^2\Big(\frac{s}{m^2}\Big)$.
This limit can be the case of the $b$ loop, its contribution can be
large in MSSM. The nature of these new DL has been clarified
in \cite{FKM,KoYa}.  It was shown that the logarithms are
related to Sudakov form factor entered into the one loop triangle
diagram. In the ref. \cite{KoYa}, we have resummed these DL explicitly. 
The result is 
\begin{eqnarray}
F&=&F^{\mbox{t-loop}}\sum\limits_{n=0}^{\infty}\frac{2\Gamma (n+1)}
{\Gamma (2n+3)}\Big( -\rho \Big)^n \to_{\rho\to\infty} 
\Big( \frac{2\ln (2\rho)}{\rho} \Big)F^{\mbox{t-loop}}.
\end{eqnarray}
First two terms in the series do 
coincide with the result of direct loop calculations.
The scale of the QCD coupling in eq.(1) should be $\mu^2\approx 9 m^2$
\cite{MeStHqq,AkWaYa1}.
The single logarithms in this sum have been resummed 
in \cite{AkWaYa1}, the result is given by the eq.(6). 

Next, once we understand that QCD corrections are small for the 
H-$\gamma\gamma$ vertex, we may ask: how large are electro-weak
corrections? 
The explicit calculation of EW corrections does not exist up to now. 
It is an extremely complicated problem. 
The only thing that we know is the leading term in the limit of the
heavy Higgs masses, 
$O(G_F m^2_H)$. This result can be taken as an estimation of the whole 
EW correction.  The calculations of the $O(G_F m^2_H)$ term has been
performed in \cite{MeYa2}.   It was suggested to use {\em the equivalence theorem}, which
states that at large Higgs masses the EW part of the Standard Model 
is described by the following $U(1)$ gauged sigma model: 
\begin{eqnarray}
L=L_0-\frac{m_H^2}{4 v^2}(\pi^2+H^2)^2-\frac{m_H^2}{v}(\pi^2+H^2)H,
\end{eqnarray}
here $H, m_H, v$ are the Higgs field, the mass and the vacuum expectation
value, $\pi =(w^+,z,w^-)$ is the triplet of the Goldstone bosons, and 
$L_0=(D_\mu w)(D^\mu w)^*+\frac{1}{2}\partial_\mu z\partial^\mu z +
\frac{1}{2}\partial_\mu H\partial^\mu H -\frac{1}{2}m_H H^2-\frac12 F^2$
The result for the Higgs-two photon coupling at two loops reads
(the details on the calculations  can be found in \cite{MeYa2})
\begin{eqnarray}
F=F^{1-loop}\Big(1-3.027\Big(\frac{m_H}{4\pi v}\Big)^2 \Big).
\end{eqnarray}
The correction is very small, less than one percent at $m_H<150$ GeV.
\section{Heavy quark production and DL} 
In the intermediate mass range, the main production process of the Higgs
boson is $\gamma\gamma\to H \to b\bar b$. 
Here we would like to consider main  background
process $\gamma \gamma \to b \bar b,$ which gets 
large QCD corrections and can spoil measurements of Higgs
properties.

 The Born order cross section for the $J_z=0$ channel of 
$\gamma \gamma \to b \bar b$
is suppressed by $\frac{m_b^2}{s}$ in comparison with 
$J_z=\pm 2$ \cite{Borden,Jikia}. 
By using the polarization of initial photons we can suppress the
background process. 
However, the perturbative QCD correction to $\gamma \gamma \to b \bar b$
in $J_z=0$ channel contains large double logarithms of the form  
$\rho=\alpha_s\ln^2(\frac{m_b^2}{s})$
at $|s|,|t|,|u| \gg m^2$.  They give contributions to the cross section 
of the same order as Born contribution. This shows the 
importance of accounting 
for large logarithmic (DL/SL) terms explicitly in all higher orders.  
The presence of the large correction was noticed by Jikia 
in \cite{Jikia}. The double logarithmic nature and the origin of these corrections 
were studied in the pivotal paper in this subject \cite{FKM} by Fadin,
Khoze and Martin. The authors studied the process with one and two loop accuracy.    
They demonstrated that the key fact which is relevant for DL analysis 
is the ``triangle topology'' of the box diagram. In more details: 
only the box diagram gives DL, moreover, the only momenta configurations
which are important are those in which one of the propagator is much
harder than others.
As a bottom line, we get four effective triangle diagrams from the box 
diagram. 
Only these effective diagrams can give DL at higher orders as well.
 
Recently, large logarithm resummation have been considered by two groups. 
First, DL resummation and RG improvement have been
addressed in the excellent set of the papers by Melles, Stirling and Khoze
\cite{MeSt1,MeSt2,MeSt3,MeSt4} (also see references therein).   
Second, DL/SL resummation with next-to-leading-log (NLL) accuracy as well
the origin and nature of the cancellations
of many diagrams at high orders have been studied by us \cite{AkWaYa,AkWaYa1}.
The authors of \cite{MeSt1} stated that the double logarithms have 
``non-Sudakov'' origin. We think that the DL in present problem are 
closely related to Sudakov DL.
Our approach is based on two facts. First, only triangle topologies of
the one loop box diagram give DL at higher loops. Second,   
the only origin of the double logarithms is the off-shell Sudakov form
factor \cite{Poggio} 
\begin{eqnarray}
S(p_1 ,p_2)=\mbox{Exp}\Big( -\frac{C_F\alpha_s(\nu^2)}{2\pi}
\ln(\frac{s}{ |p_1|^2})\ln(\frac{s}{ |p_2|^2}) \Big),
\quad\mbox{with $m^2 \ll |p_1|^2, |p_2|^2 \ll s$}.
\end{eqnarray}
This has to be included in all triangle topologies (fig.1) 
of the one-loop box diagram.  
But more importantly, we have proved that the rest of the high loop
diagrams (other than those accounted for in eq.(4)) will 
either cancel in the subgroups of the diagrams
or develop standard on-shell Sudakov exponent due to final $b\bar b$ lines.
We refer reader to \cite{AkWaYa} for details. The result for the
amplitude is the sum of three triangle topologies (A,B,$C_1=C_2$ in fig.1):
$F=F_A+F_B+2F_C$. The topology A gives a standard on-shell Sudakov form factor.
The result for the B topology reads
\begin{eqnarray}
F^{DL}_B&=&F_B^{\mbox{1-loop}}\sum\limits_{n=0}^{\infty}
\frac{2\Gamma (n+1)}{\Gamma (2n+3)}
\Big( -\rho_B \Big)^n=_2F_{2}(1,1;2,\frac{3}{2},-\frac{\rho_B}{4})
F_B^{\mbox{1-loop}},
\end{eqnarray}
with $\rho_B=\frac{C_F\alpha_s(\mu^2)}{2\pi}L^2, L=
\ln\Big(\frac{m^2}{s}\Big)$. 
The topology C differs from B only in color structure. 
The answers for B and C topologies in the DL approximation 
are related by simple substitution: $C_F\to C_A/2.$  
 It is possible to develop this approach to achieve
next-to-leading-logarithmic accuracy \cite{AkWaYa}. 
 The final result for the next-to-leading-logarithmic form factor 
for the topology B (extra to (5)) reads \cite{AkWaYa}
\begin{eqnarray}
F_B^{NLL}&=&\frac{1}{L}
F^{1-loop}_B\sum_{n=0}^{\infty} \frac{\Gamma (n+1)}{\Gamma (2n+2)}
(-\rho_B )^n\Big(3-\frac{\rho_B\beta_0}{C_F}\frac{n}{2n+2}\Big( 
\frac{n+1}{2n+3}+\frac{\ln (s/\mu^2)}{L}\Big) \Big),   
\end{eqnarray} 
with $\beta_0=11-\frac{2n_f}3$. The relative size of SL corrections 
in comparison to the DL contributions is estimated as $\frac{1}{L}$, 
which is of order 30\%. The normalization point in (4) is
$\nu^2=\frac{|p_1^2p_2^2|}{s}$, which corresponds to $\mu^2\approx 9 m^2$
in (5,6) \cite{AkWaYa,MeSt3}.
More details on the topology C with NLL accuracy can be found in
\cite{AkWaYa}, and on numerics in \cite{AkWaYa,MeSt4}.\\
\begin{figure}
\psfig{figure = 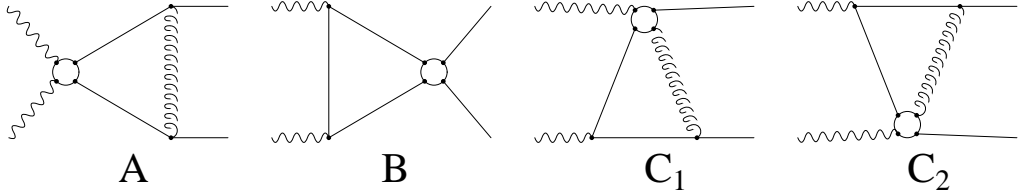,height=1.0in}
\caption{\label{fig1}
The triangle topologies of the one loop box diagram.}
\end{figure}
{\bf Acknowledgements:} The work is supported by the 
US Department of Energy. I would like to thank R. Akhoury, H. Wang, 
K. Melnikov, J. Koerner, M. Kotsky for their collaborations on the  
results presented here. 


\begin{thebibliography}{9}
\bibitem{Higgs} P. W. Higgs, Phys. Rev. Lett 12, 132 (1964),
Phys. Rev. 145 1156 (1966).
\bibitem{Kane} J.F. Gunion, H.E. Haber, G. Kane, and S. Dawson,
The Higgs Hunter's Guide (Addison-Wesley, Reading, MA, 1990).
\bibitem{LEPHIGGS} A. Read, LEPC presentation on July 2000, 
http://lephiggs.web.cern.ch/LEPHIGGS.
\bibitem{MeYa1} K.~Melnikov and O.~Yakovlev,
Phys.\ Lett.\  {\bf B324}, 217 (1994).
\bibitem{MeYa2} J.~G.~Koerner, K.~Melnikov and O.~Yakovlev,
Phys.\ Rev.\  {\bf D53}, 3737 (1996).
\bibitem{ZeSp1}
A.~Djouadi, M.~Spira and P.~M.~Zerwas,
Phys.\ Lett.\  {\bf B311}, 255 (1993).
\bibitem{KoYa}
M.~I.~Kotsky and O.~I.~Yakovlev,
Phys.\ Lett.\  {\bf B418}, 335 (1998). 
\bibitem{MeStHqq}
M.~Melles,
Phys.\ Rev.\  {\bf D60}, 075009 (1999).
\bibitem{Jikia} G. Jikia, A. Takabladze, Phys. Rev. D54, 2030 (1996).
\bibitem{FKM} V.~S.~Fadin, V.~A.~Khoze and A.~D.~Martin,
Phys.\ Rev.\  {\bf D56}, 484 (1997).
\bibitem{AkWaYa} R. Akhoury, H. Wang, O. Yakovlev, 
``On large logarithms in $\gamma\gamma \to b\bar b$ '', MCTP-09.
\bibitem{AkWaYa1} R. Akhoury, H. Wang, O. Yakovlev, 
``On large logarithms in $H\to \gamma\gamma $ '', MCTP-11.
\bibitem{Borden} D.L. Borden et al, Phys. Rev. D50 , 4499 (1994).
\bibitem{MeSt1} M. Melles, W.J. Stirling, Phys. Rev. D59, 
094009 (1999).
\bibitem{MeSt2} M.~Melles and W.~J.~Stirling,
Eur.\ Phys.\ J.\  {\bf C9}, 101 (1999).
\bibitem{MeSt3}
M.~Melles and W.~J.~Stirling,
Nucl.\ Phys.\  {\bf B564}, 325 (2000).
\bibitem{MeSt4}
M.~Melles, W.~J.~Stirling and V.~A.~Khoze,
Phys.\ Rev.\  {\bf D61}, 054015 (2000).
\bibitem{Poggio} J. Carazzone, E. C. Poggio, H. R. Quinn, Phys. Lett. B
57, 161 (1975).
\end{thebibliography}
\end{document}